\documentclass{aipproc}
\usepackage{graphicx}
\usepackage{amssymb}
\usepackage{graphicx}
\usepackage{subfigure}

\layoutstyle{6x9}
\usepackage{fixltx2e}
\begin{document}

\title{Using temporal distributions of transient events to characterize cosmological source populations}
 \vspace{-3mm}

\author{ E.~Howell, D. Coward, R. Burman and D. Blair}{
  address={School of Physics, the University of Western Australia,\\
 35 Stirling Hwy, Crawley, WA 6009, Australia, \\ ejhowell@physics.uwa.edu.au}
}

\begin{abstract}
The brightest events in a time series of cosmological transients obey an observation time dependence which is often overlooked. This dependence can be exploited to probe the global properties of electromagnetic and gravitational wave transients \cite{ejh2, cow1}. We describe a new relation based on a peak flux--observation time distribution and show that it is invariant to the luminosity distribution of the sources \cite{ejh1}. Applying this relation, in combination with a new data analysis filter, to \emph{Swift} gamma-ray burst data, we demonstrate that it can constrain their rate density.
\end{abstract}

\maketitle

\graphicspath{{./}{Figures/}}

\section{Introduction}
 \vspace{-3mm}
Transient astronomy is a rapidly emerging field which will benefit enormously as coordinated observations of sources emitting electromagnetic photons, gravitational waves, cosmic rays or neutrinos become commonplace. Transient sources, with peak emission durations much less than typical observational periods, will include events such as supernovae, gamma-ray bursts (GRBs) and gravitational wave emissions from compact object inspirals \citep{ejh2}.

For cosmological populations of sources, the brightness distribution is conventionally used to constrain their luminosity function (LF), their evolution in density \citep{Peebles} and for transient sources, their rate density
\citep{schmidt01}. Estimates of global properties are obtained by fitting the number --
brightness distribution to models that include luminosity, source density and evolution effects. For the case of
transient events an additional parameter is available -- the event arrival times.

\begin{figure}
\includegraphics[scale=0.68,bbllx = 0pt,bblly =337pt, bburx = 559 pt, bbury = 503 pt,origin=lr]{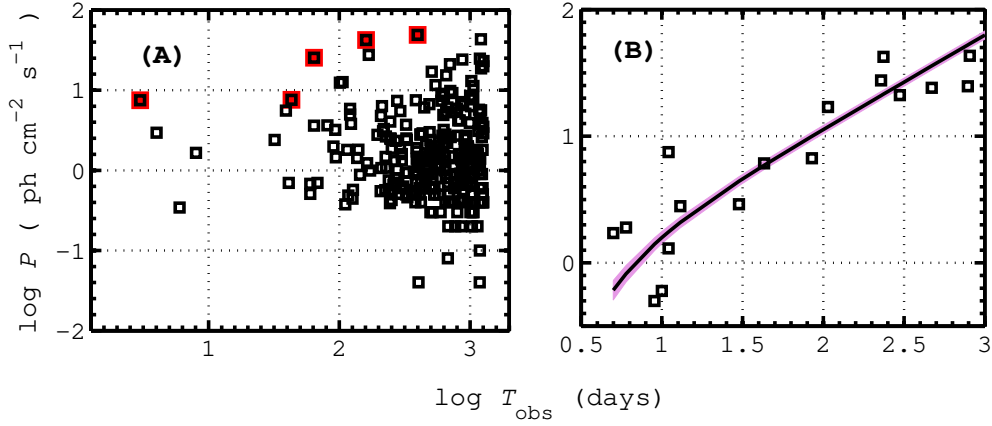}
\caption{(A) The observation time - peak flux distribution of long gamma-ray bursts as observed by the \emph{Swift} satellite. Successively brighter events that have occurred during this period are highlighted. (B) PEH fits to the \emph{Swift} LGRB data sample. The estimates
 are consistent with estimates by \cite{guetta07} of $\hspace{0.1cm} 0.10^{+0.08}_{-0.06}\hspace{0.1cm}\mathrm{Gpc}^{-3}\mathrm{yr}^{-1}$ using the same model parameters. }
\label{fig_swift_PeakFluxTriangle}
\end{figure}

The brightest events of an astrophysical transient population obey a simple, not well known, observation-time dependence. Figure \ref{fig_swift_PeakFluxTriangle}(A) illustrates this dependence by way of the observation time - peak flux distribution of long duration ($\geq 2\hspace{0.5mm}$s GRBs (LGRBs) as observed by the \emph{Swift} satellite\footnote{http://swift.gsfc.nasa.gov/docs/swift/swiftsc.html}. Although GRBs are not standard candles and observations of these transients suffer from selection effects, two things are immediately apparent from this plot: Firstly, the initial observations are of the intermediate energy sources -- these events occur frequently and correspond to the high-$z$ peak of the event distribution. Secondly, as observation time increases, rarer events from both $z$ extremes of the distribution come into play, producing a triangular distribution in log-log space. The events that are missed through observational bias or background detector noise will not generally be amongst the brightest events in a distribution. Therefore, to extract information about the underlying source population it could be beneficial to use the the brightest events in a survey.

The temporal distribution of transient astrophysical populations of events has been described by the `probability event
horizon' (PEH) concept of \cite{cow1}. This method establishes a temporal dependence by noting the occurrences of
successively brighter events in a time series -- shown by the highlighted events in Figure \ref{fig_swift_PeakFluxTriangle}(A). By utilizing the fact that the rarest events will preferentially occur
after the longest observational periods, it produces a data set with a unique statistical signature. To extract this signature from a body of data, \citet{ejh2}, developed a PEH filter, which used in unison with a brightness -- observation time relation, could be applied to obtain estimates of the rate density of events, $r_{0}$, from a standard candle population of sources. In its simplest form, the PEH filter simply extracts successively brighter events as a function of observation time. However, for the case of GRBs and many other transient sources, there is a distribution in luminosities. This imposes a severe restriction on the technique. To overcome this problem, \citet{ejh1} derived a new \emph{peak flux -- observation time} relation that can be applied to distributions of transient astrophysical events and is independent of the form of the LF. This relation will be described in the next section.

\section{The peak flux -- observation time relation}
\label{section_derivation}
 \vspace{-3mm}

Consider a distribution of events distributed in a Euclidean space by an event rate density $r_{0}$ and a LF $\phi(L)$ $(L_{\mathrm{min}}\le L \le L_{\mathrm{max}})$.
The observed peak flux, $P$ (photons
$\mathrm{cm}^{-2}\mathrm{s}^{-1}$), or `brightness', distribution of events over an observation time $T$ is a convolution of the
radial distribution of the sources and their LF. For $r_{0}$ and $\phi(L)$ independent of position,
setting the upper limit of distance $r$ is the maximum range for which an event with luminosity $L$, produces a peak flux
$P$. Integrating over the radial distance yields:

\begin{equation}
N(> P) = \frac{T r_{0}\hspace{0.5mm} \Delta\Omega/4\pi }{3 \hspace{0.5mm}\sqrt{4 \pi}}\hspace{0.5mm}
P^{-3/2}\hspace{0.5mm}\int_{L_{\mathrm{min}} }^{L_{\mathrm{max}} } \phi(L)L^{3/2} dL\,,
 \label{eq_NF2}
\end{equation}

\noindent where the average solid angle covered on the sky has been accounted for by $\Delta\Omega /4 \pi$. This is the familiar log $N$--log $P$ relation, \mbox{$N(>P) \propto P^{-3/2}$}, a power law independent of
the form of the LF \citep{horack94}.

To see how the events are distributed in time, we note that, as the
individual events will follow a Poisson distribution in time the temporal separation between events follow an exponential distribution, defined by a mean event rate $R(r) = r_{0}(4/3)\pi r^{3}$ for events out to $r$.
The probability for at least one event $>P$ to occur in a volume bounded by $r$ during an observation time $T$ at
constant probability $\epsilon$ is given by:

\vspace{-3mm}
\begin{equation}
\label{eq_peh}
\mathcal{P}(n \ge 1;R(r),T)=  1 - e^{R(r)T} = \epsilon \,.
\end{equation}

\noindent For this equation to remain satisfied with increasing observation time:

\vspace{-3mm}
\begin{equation}\label{eq_eps}
N(>P) = R(r)T =  |\mathrm{ln}(1 - \epsilon)|. \\
\end{equation}

Equations (\ref{eq_NF2}) and (\ref{eq_eps}) for $N(>P)$ combine to give the relation for the evolution of brightness as
a function of observation time \cite{ejh1}:

\begin{equation}
P(T)\! =\!\! \left(\frac{r_{0}\hspace{0.5mm}\Delta\Omega /4\pi\hspace{0.5mm} }{3 \hspace{0.5mm}\sqrt{4
\pi}\hspace{0.5mm}|\mathrm{ln}(1 - \epsilon)|}\right)^{\!\!2/3}\!\! \left[\int_{L_{\mathrm{min}}
}^{L_{\mathrm{max}}}\!\!\!\!\!\!\!\phi(L)L^{3/2}dL \right]^{2/3}\!\!\!\!\!\!T^{2/3}\!\hspace{0.5mm}.
\label{eq_logPlogT_euc}\\
\end{equation}

\vspace{2mm}
This relation shows that for a Euclidean event distribution, a log $P$\hspace{0.4mm}--\hspace{0.2mm}log $T$ distribution will have a slope of 2/3,
independent of the form of the LF. We can use the log $P$\hspace{0.4mm}--\hspace{0.5mm}log $T$ relation to produce curves defining the probability, $\epsilon$, of obtaining some
value of peak flux, $P$, within an observation-time, $T$, for a given $r_{0}$ \mbox{and $\phi(L)$}.

\section{A time-based signal processing filter}
 \vspace{-3mm}

Figure \ref{fig_swift_PeakFluxTriangle}(A) shows that in its simplest form, the PEH filtered output consists of only a small sub-sample of the overall population. Thus, estimates of parameters obtained by fitting to a model PEH curve will inevitably suffer from low resolution. To increase the data sample, one can split a time series into smaller time segments, apply the filter to each and recombine the output \citep{ejh2}. The use of this technique is subtle, as splitting can increase the number of outlying events and can diminish the statistical signature of the output. In addition, there is a significant probability of a bright event occurring early in an observational period, further reducing the sample size.

To increase the amount of usable data and to enhance the signature of the filter output, two physical principles can be exploited \cite{ejh1}. Firstly, the \emph{temporal cosmological principle}, implies that for time scales that are short compared to the age of the Universe, there is nothing special about the time we make an observation. Thus, the PEH signature of a transient population of events is independent of when a detector is switched on.
\emph{Time reversal invariance} means that a time series of events can be treated as a closed loop. Therefore, the PEH filter can be applied to a data set from any start time and in both temporal directions. As the start time for the PEH analysis is now arbitrary, the filter can be applied in such a way that the brightest event can be set as the final event in a series, and the observational period will be the total length of the loop. These techniques can increase the PEH sample and significantly improve the statistical signature of the filter output.

\section{Application to \emph{SWIFT} data}
 \vspace{-3mm}
To demonstrate the log $P$\hspace{0.5mm}--\hspace{1.5mm}log $T$ technique, \citet{ejh1} used the LGRB peak flux data recorded by the \emph{Swift} satellite between 2004 December and 2007 April. Validation was achieved by fitting a 90\% confidence log $P$\hspace{0.5mm}--\hspace{1.5mm}log $T$ threshold to PEH filtered data, using recent estimates of source parameters from \cite{guetta07}.

In this section, we explore the efficiency of the method as a practical tool for probing global source parameters, using a least-squares fit to the same data. Estimates of the parameters are compared with those of \cite{guetta07}, who fitted to the brightness distribution to estimate the rate density and LF parameters of the LGRB population.

To allow for a cosmological distribution of sources, equation (\ref{eq_logPlogT_euc}) must be modified to allow for cosmic
evolution. A standard Friedman cosmology is used to define a differential event rate, $\mathrm{d}R(z)$, in the
redshift shell $z$ to $z + \mathrm{d}z$. The luminosity and flux will be related through $z$ by a luminosity distance
$d_{\mathrm{L}}^{2}(z)$ \citep[see, e.g.,][]{cow1,pm01}. In this case, solving equation
(\ref{eq_eps}) numerically, with $P = L/4\pi d_{\mathrm{L}}^{2}(z)$, will \hspace{0.5mm}yield \hspace{0.5mm}the \hspace{0.5mm}cosmological \mbox{\hspace{0.5mm}log $P$\hspace{0.5mm}--\hspace{1.5mm}log $T$}
relation. We apply this fit to the same GRB data, setting $\epsilon$ to $ 50\%$, the mean value obtained from 10,000 simulated data sets.

Figure \ref{fig_swift_PeakFluxTriangle}(B) shows the log $P$\hspace{0.5mm}--\hspace{1.5mm}log $T$ fit to the \emph{Swift} LGRB sample, yielding
$r_{0} = \hspace{0.12cm} 0.10\pm 0.02\hspace{0.1cm}\mathrm{Gpc}^{-3}\mathrm{yr}^{-1}$ at a $1\hspace{0.2mm}$--$\hspace{0.2mm}\sigma$ confidence level -- this is within the range of $r_{0} = \hspace{0.1cm} 0.10^{+0.08}_{-0.06}\hspace{0.1cm}\mathrm{Gpc}^{-3}\mathrm{yr}^{-1}$ obtained by \cite{guetta07} using four free parameters. The greater resolution of the log $P$\hspace{0.5mm}--\hspace{1.5mm}log $T$ fit is a result of using only one free parameter, $r_{0}$.

The log $P$\hspace{0.5mm}--\hspace{1.5mm}log $T$ fit utilizes the brightest events in a data stream, which on average correspond to the closest events \citep{bagoly06}. Therefore, an accurate representation of the high-$z$ selection function may not be essential. Additionally, the technique may be a useful probe of the high end of the LF. Preliminary studies \citep{Howell_2010} suggest the method may shed some light on the narrow observed distribution of BATSE peak energies \citep{Cohen_batse_bias_98, Piran_BATSEhighEGRB_95}. Finally, we note that the time dependence conveniently allows the method to be used as a predictive tool, to determine the probability of an event with some predefined energy to occur within some observation time \cite{ejh1,coward05b}.

 \vspace{-4mm}

\end{document}